\documentclass[twocolumn,showpacs,preprintnumbers,amsmath,amssymb]{revtex4}

\usepackage{graphicx}
\usepackage{dcolumn}
\usepackage{bm}

\begin{document}

\preprint{}
\input{epsf.tex}

\epsfverbosetrue

\title{Lifetime and Emission Characteristics of Electronic Excitations in 2D Optical Lattices}

\author{Hashem Zoubi, and Helmut Ritsch}

\affiliation{Institut fur Theoretische Physik, Universitat Innsbruck, Technikerstrasse 25, A-6020 Innsbruck, Austria}

\date{15 March, 2011}

\begin{abstract}
Collective electronic excitations ``excitons'' in planar optical lattices exhibit  strong modifications of the radiative damping rate and directional emission pattern as compared to a single excited atom. Excitons for long wave numbers and polarizations orthogonal to the lattice plane exhibit superradiance with a very short life time and a tightly confined emission direction. For shorter wavelength and in plane polarization they can  posses a long life time, which  beyond a critical wave number tends to infinity. Those excitons thus become metastable and decoupled from the free radiation field. They can store a single energy quantum  for a long time and transfer the excitation over long distances. In general the spatial, polarization and frequency dependence of the emission pattern can provide us with optical and electronic properties of optical lattices.
\end{abstract}


\maketitle

Optical lattices with  ultracold atoms constitute an established test system for a wide range of condensed matter models and phenomena \cite{Dalibard,Lewenstein}. Recent theoretical and experimental progress highlights the importance of such a system both for fundamental physics and applications. Ultracold atoms loaded to optical lattices are well described by the Bose-Hubbard model \cite{Jaksch} that predicts the quantum phase transition from the superfluid into the Mott insulator phase \cite{Greiner}. The big similarity between optical lattice ultracold atoms in the Mott insulator phase and solid crystals encouraged us to study different solid state effects for such a set-up. Many advantages of optical lattices over solid crystals are a result of the precise controllability of the system properties, e.g. the number of atoms per site, the geometry and symmetry. These advantages opened the door for a deep understanding of different solid state effects and to answer open questions in the field, beside the emergence of new phenomena \cite{Dalibard}.

At very low temperature in optical lattices the atoms are localized in the lowest Bloch band forming a Mott insulator phase. The atoms retain their identity except from light shifts due to the external laser fields that form the optical lattice \cite{Jaksch,Greiner}. This fact makes optical lattices analogous to molecular crystals. Optical properties of molecular crystals are mainly dominated by the formation of collective electronic excitations, in which a single molecule electronic excitation is delocalized in the crystal due to electrostatic interactions and represented by a wave that propagates in the material with a fixed quasi-momentum and effective mass. Such quasi-particles are termed as Frenkel excitons \cite{Davydov}. Excitons in optical lattice ultracold atoms studied extensively by us \cite{ZoubiA,ZoubiB}.

In the present letter we investigate the radiative damping rate and emission pattern of electronic excitations in planar optical lattices. Optical excitations in an ultracold gas are short lived and lead to strong heating and a fast decay of coherence via emission recoil and reabsorption, hence most experiments are performed very far from resonance. The formation of excitons in 2D optical lattices can significantly alter the radiative damping rate into free space and its impact on the lattice atoms. We show that excitons will decay collectively in superradiant or subradiant fashion, and can be even metastable. As the decay and absorption occur collectively, the recoil and energy difference are spread over large spatial regions, depositing only a small amount of energy and momentum per particle. A 2D Mott insulator thus can only be excited under special conditions and will remain robust under near resonant excitations.  

Historically the modified radiative damping rate of excitons at low dimensional molecular crystals was predicted by Agranovich \cite{Agranovich}, where superradiant decay foreseen for small wave vector excitons. While in those cases experimental confirmation are difficult and superimposed by many other effects, the situation in cold atom optical lattices is much more promising. The special case of 1D optical lattice was treated by us in \cite{ZoubiC}. Experimentally the fabrication of 2D optical lattices in the Mott insulator phase was achieved experimentally by several groups \cite{Spielman}. Recently the fluorescence imaging of 2D optical lattice in the Mott insulator phase is obtained down to a resolution of a single site \cite{Sherson}. The emission pattern for long wavelength excitons we derive here can serve as an observation tool for the kind of excitations exist in the system, and to predict the formation of excitons, and allow us to extract different properties of the system.

\ 

As our model system we consider a two-dimensional planar optical lattice in the Mott insulator phase with one atom per site. The optical lattice has a square symmetry of lattice constant $a$, see figure (1). The atoms are taken to be two-level systems of transition energy $E_A$. An electronic excitation can transfer among atoms at different sites due to electrostatic interactions with the coupling parameter $J_{\bf nm}$, where ${\bf n}$ is the site position in the lattice plane. Here ${\bf n}=(n_x,n_y)$ with $n_i=a(0,\pm 1,\cdots,\pm N_i/2)$, and $N=N_x\times N_y$ is the number of lattice sites. The electronic excitations are described by the Hamiltonian \cite{ZoubiA}
\begin{equation}
H_{ex}=\sum_{\bf n}E_A\ B_{\bf n}^{\dagger}B_{\bf n}+\sum_{\bf nm}J_{\bf nm}\ B_{\bf n}^{\dagger}B_{\bf m}.
\end{equation}
$B_{\bf n}^{\dagger}$ and $B_{\bf n}$ are the creation and annihilation operators of an electronic excitation at site ${\bf n}$, respectively. The operators are assumed to obey Bose commutation relations, which is a good approximation at low density of electronic excitations in the lattice. The transition dipole operator of an electronic excitation is $\hat{\mbox{\boldmath$\mu$}}=\mbox{\boldmath$\mu$}\sum_{\bf n}\left(B_{\bf n}+B_{\bf n}^{\dagger}\right)$.

An electronic excitation is delocalized in the lattice due to, e.g., resonance dipole-dipole interactions, and in exploiting the lattice symmetry it represented by a wave that propagates in the lattice with in-plane wave vector ${\bf k}$. Such collective electronic excitations are called excitons. In using the Fourier transform $B_{\bf n}=\frac{1}{\sqrt{N}}\sum_{\bf k}e^{i{\bf k}\cdot{\bf n}}B_{\bf k}$, we get the diagonal exciton Hamiltonian $H_{ex}=\sum_kE_{ex}({\bf k})\ B_{\bf k}^{\dagger}B_{\bf k}$, with the energy dispersion $E_{ex}({\bf k})=E_A+\sum_{\bf R}J({\bf R})e^{i{\bf k}\cdot{\bf R}}$. The coupling parameter is a function of the distance between the sites, namely ${\bf R}={\bf n-m}$. For the case of nearest neighbor interactions with the coupling parameter $J$, we get $E_{ex}({\bf k})=E_A-2J\left[\cos(k_xa)+\cos(k_ya)\right]$. The wave numbers take the values $k_i=\frac{2\pi}{N_ia}\left(0,\pm 1,\cdots,\pm N_i/2\right)$.

The optical lattice is located in free space. The radiation field of free space is represented by the Hamiltonian $H_{rad}=\sum_{{\bf q}\lambda}E_{ph}(q)\ a_{{\bf q}\lambda}^{\dagger}a_{{\bf q}\lambda}$, where $a_{{\bf q}\lambda}^{\dagger}$ and $a_{{\bf q}\lambda}$ are the creation and annihilation operators of a photon with wave vector ${\bf q}$ and polarization mode $\lambda$, respectively. The photon dispersion has the linear dispersion $E_{ph}(q)=\hbar\omega_r(q)=\hbar cq$. The electric field operator is
\begin{equation}
\hat{\bf E}({\bf r})=i\sum_{{\bf q}\lambda}\sqrt{\frac{\hbar cq}{2\epsilon_0 V}}\left\{a_{{\bf q}\lambda}\ {\bf e}_{{\bf q}\lambda}e^{i{\bf q}\cdot{\bf r}}-a_{{\bf q}\lambda}^{\dagger}\ {\bf e}_{{\bf q}\lambda}^{\ast}e^{-i{\bf q}\cdot{\bf r}}\right\},
\end{equation}
where ${\bf e}_{{\bf q}\lambda}$ is the photon polarization unit vector, and $V$ is the normalization volume.

The matter-field coupling is given by the electric dipole interaction $H_I=-\hat{\mbox{\boldmath$\mu$}}\cdot\hat{\bf E}$, and in the rotating wave approximation for linear polarizations we get
\begin{eqnarray}
H_I&=&-i\sum_{{\bf q}\lambda,{\bf n}}\sqrt{\frac{\hbar cq}{2\epsilon_0 V}}\left(\mbox{\boldmath$\mu$}\cdot{\bf e}_{{\bf q}\lambda}\right) \nonumber \\
&\times&\left\{a_{{\bf q}\lambda}B_{\bf n}^{\dagger}\ e^{i{\bf q}\cdot{\bf n}}-a_{{\bf q}\lambda}^{\dagger}B_{\bf n}\ e^{-i{\bf q}\cdot{\bf n}}\right\}.
\end{eqnarray}
We define ${\bf q}=({\bf q}',q_z)$, in using the above transformation and the relation $\frac{1}{N}\sum_{\bf n}e^{i({\bf q}'-{\bf k})\cdot{\bf n}}=\delta_{{\bf q}'{\bf k}}$, we obtain
\begin{equation}
H_I=\sum_{{\bf q}\lambda}i\hbar g_{{\bf q}\lambda}\left\{b_{{\bf q}\lambda}B_{\bf k}^{\dagger}-b_{{\bf q}\lambda}^{\dagger}B_{\bf k}\right\},
\end{equation}
with the coupling parameter $\hbar g_{{\bf q}\lambda}=-\sqrt{\frac{\hbar cqN}{2\epsilon_0 V}}\left(\mbox{\boldmath$\mu$}\cdot{\bf e}_{{\bf q}\lambda}\right)$, where now ${\bf q}=({\bf k},q_z)$, and $q^2=k^2+q_z^2$, see figure (1). The coupling is between photons with wave vector component parallel to the lattice equal to the exciton in-plane wave vector.

\begin{figure}[h!]
\centerline{\epsfxsize=5cm \epsfbox{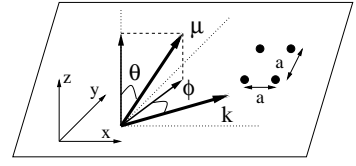}}
\caption{The optical lattice plane, with lattice constant $a$. The directions of the in-plane wave vector ${\bf k}$ and the transition dipole \boldmath$\mu$ are seen.}
\end{figure}

Here we calculate the radiative damping rate into free space of a single exciton with wave vector ${\bf k}$ in a planar optical lattice. We use the Fermi Golden rule $\Gamma_k=\frac{2\pi}{\hbar}\sum_{q_z\lambda}\left|\langle f|H_I|i\rangle\right|^2\delta\left(E_{ex}({\bf k})-E_{ph}(q)\right)$. The initial state is of a single exciton and empty radiation field, $|i\rangle=|1_{ex}({\bf k}),0_{ph}\rangle$, and the final state is of a single photon and ground state optical lattice, $|f\rangle=|0_{ex},1_{ph}({\bf q})\rangle$. The matrix element is given by $\langle f|H_I|i\rangle=i\sqrt{\frac{\hbar cqN}{2\epsilon_0 V}}\left(\mbox{\boldmath$\mu$}\cdot{\bf e}_{{\bf q}\lambda}\right)$. The summation over the photon polarization is obtained by the relation $\sum_{\lambda}\left|\mbox{\boldmath$\mu$}\cdot{\bf e}_{{\bf q}\lambda}\right|^2=\left|\mbox{\boldmath$\mu$}\right|^2-\frac{\left|{\bf q}\cdot\mbox{\boldmath$\mu$}\right|^2}{q^2}$. The sum over $q_z$ is converted into the integral $\sum_{q_z}\rightarrow \frac{L}{2\pi}\int_0^{\infty}dq_z$, where $V=SL$ with $S=Na^2$. We define $\mbox{\boldmath$\mu$}=(\mbox{\boldmath$\mu$}_{\parallel},\mu_z)$, see figure (1). The angle between the in-plane transition dipole and the exciton wave number is $\phi$, namely $\left({\bf k}\cdot\mbox{\boldmath$\mu$}_{\parallel}\right)=\mu_{\parallel}k\cos\phi$. Furthermore, we define $\theta$ the angle between the transition dipole and the ${\bf z}$ direction, hence $\mu_{\parallel}=\mu\sin\theta$ and $\mu_z=\mu\cos\theta$. Straightforward integration yields
\begin{eqnarray}
&&\Gamma_{ex}({\bf k})=\frac{\mu^2}{2\epsilon_0a^2\hbar^2c}\frac{E_{ex}^2({\bf k})}{\sqrt{E_{ex}^2({\bf k})-E_{0}^2(k)}} \nonumber \\
&\times&\left\{\sin^2\theta\left(1-\cos^2\phi\frac{E_{0}^2(k)}{E_{ex}^2({\bf k})}\right)+\cos^2\theta\frac{E_{0}^2(k)}{E_{ex}^2({\bf k})}\right. \nonumber \\
&-&\left.2\sin\theta\cos\theta\cos\phi\frac{E_{0}(k)}{E_{ex}^2({\bf k})}\sqrt{E_{ex}^2({\bf k})-E_{0}^2(k)}\right\},
\end{eqnarray}
where $E_{0}(k)=\hbar ck$. The most important result here is that for $k\geq k_c$ we have $\Gamma_{\bf k}=0$, where $\hbar ck_c=E_{ex}({\bf k}_c)$, as beyond the singularity point the damping rate becomes imaginary. The results need to be compared with the radiative damping rate of a single atom in free space, which is $\Gamma_{at}=\frac{\mu^2E_A^3}{3\epsilon_0\pi\hbar^4c^3}$.

Here we use typical numbers for optical lattice ultracold atoms. The lattice constant is $a=1000\ \AA$, and the transition dipole is $\mu=1\ e\AA$. The exciton energy is taken to be $E_{ex}({\bf k})=1\ eV$, where for simplicity we neglect the ${\bf k}$ dependence, which is negligible relative to the $E_{0}(k)$ one. We plot the scaled damping rate, $\Gamma_{ex}({\bf k})/\Gamma_{at}$, as a function of different directions of the transition dipole, $\theta$, different directions of the exciton, $\phi$, and different exciton wave vectors, $E_{0}(k)=\hbar ck$.

In figure (2) we plot $\Gamma_{ex}/\Gamma_{at}$ vs. $E_{0}(k)$ for $\theta=0$, where the transition dipole has only component normal to the lattice plane. Here $\Gamma_{ex}({\bf k})$ is $\phi$ independent. The damping rate starts from zero, where the long wave length excitons are metastable, and increases with $k$ to become superradiant with damping rate larger that $\Gamma_{at}$. Close to $\hbar ck_c=1\ eV$ the damping rate diverges, and beyond $\hbar ck_c=1\ eV$ jumps back to zero, where all excitons between $k_c$ and the Brillouin zone boundary at $\pi/a$ are metastable, where $\hbar c\pi/a\sim 6.2\ eV$.

\begin{figure}[h!]
\centerline{\epsfxsize=4cm \epsfbox{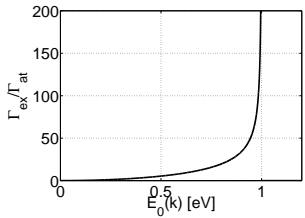}}
\caption{The scaled damping rate $\Gamma_{ex}/\Gamma_{at}$ vs. $E_{0}(k)=\hbar ck$, for $\theta=0$, and which is $\phi$ independent.}
\end{figure}

In figures (3) we plot $\Gamma_{ex}/\Gamma_{at}$ vs. $E_{0}(k)$ for $\theta=\pi/4$, where the transition dipole has equal in-plane and normal components. In figure (3a) we take the in-plane wave vector ${\bf k}$ to be parallel to the in-plane transition dipole, that is $\phi=0$. Very interesting behavior appears here for the damping rate. For small in-plane wave vectors the excitons are superradiant with damping rate larger than $\Gamma_{at}$, and decreases in increasing $k$ till it becomes zero at $E_0(k)=E_{ex}({\bf k})/\sqrt{2}$ where excitons around this point are metastable. For larger $k$ the damping rate start to increase and diverges at $E_0(k_c)$, and then jump back to zero beyond this point. In figure (3b) we take the in-plane wave vector ${\bf k}$ to be normal to the in-plane transition dipole, that is $\phi=\pi/2$. Now the small wave vector excitons are superradiant and the damping rate increase with $k$ to diverge at $k_c$, and then jump to zero.

\begin{figure}[h!]
\centerline{\epsfxsize=4cm \epsfbox{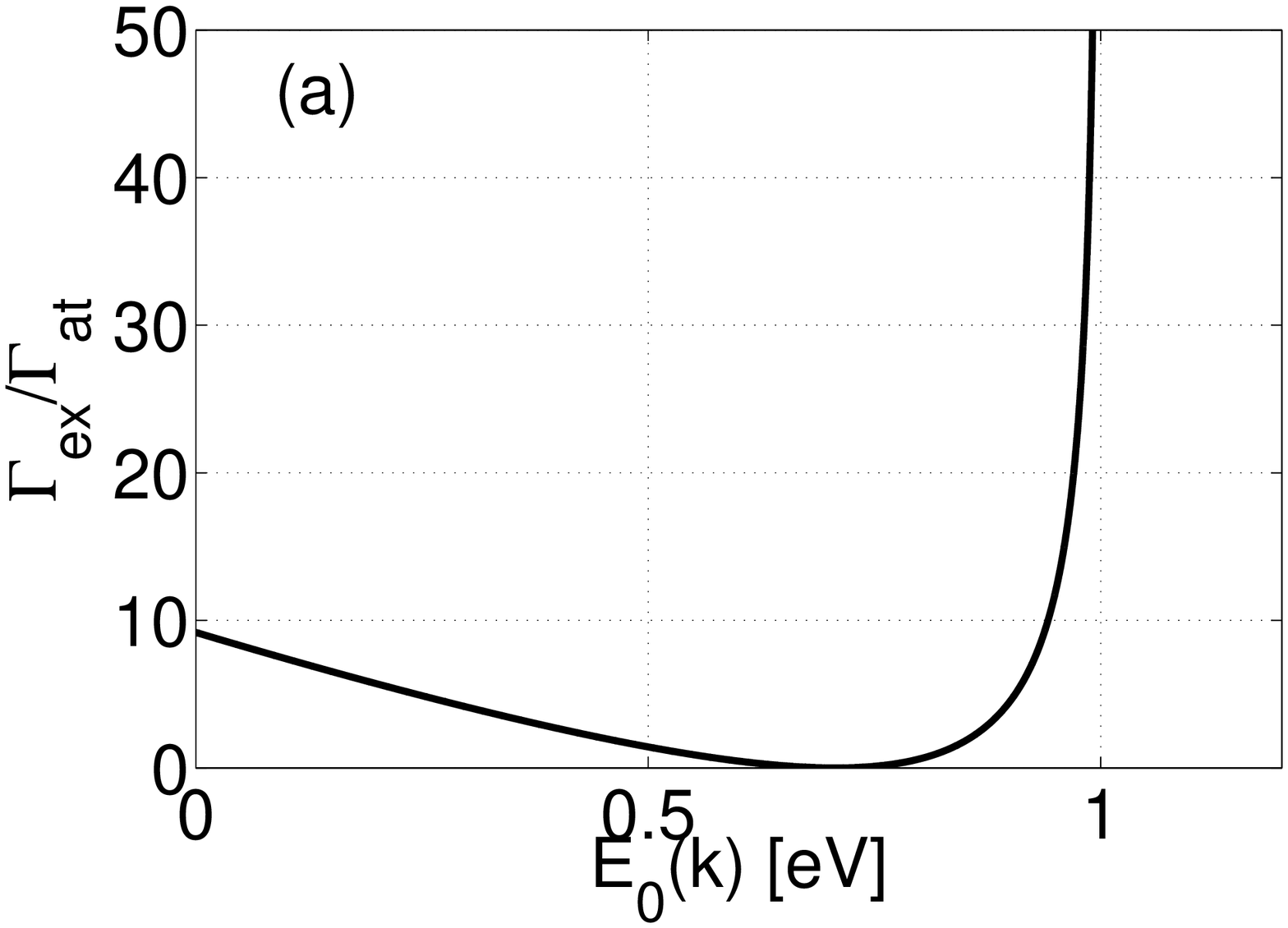}\epsfxsize=4cm \epsfbox{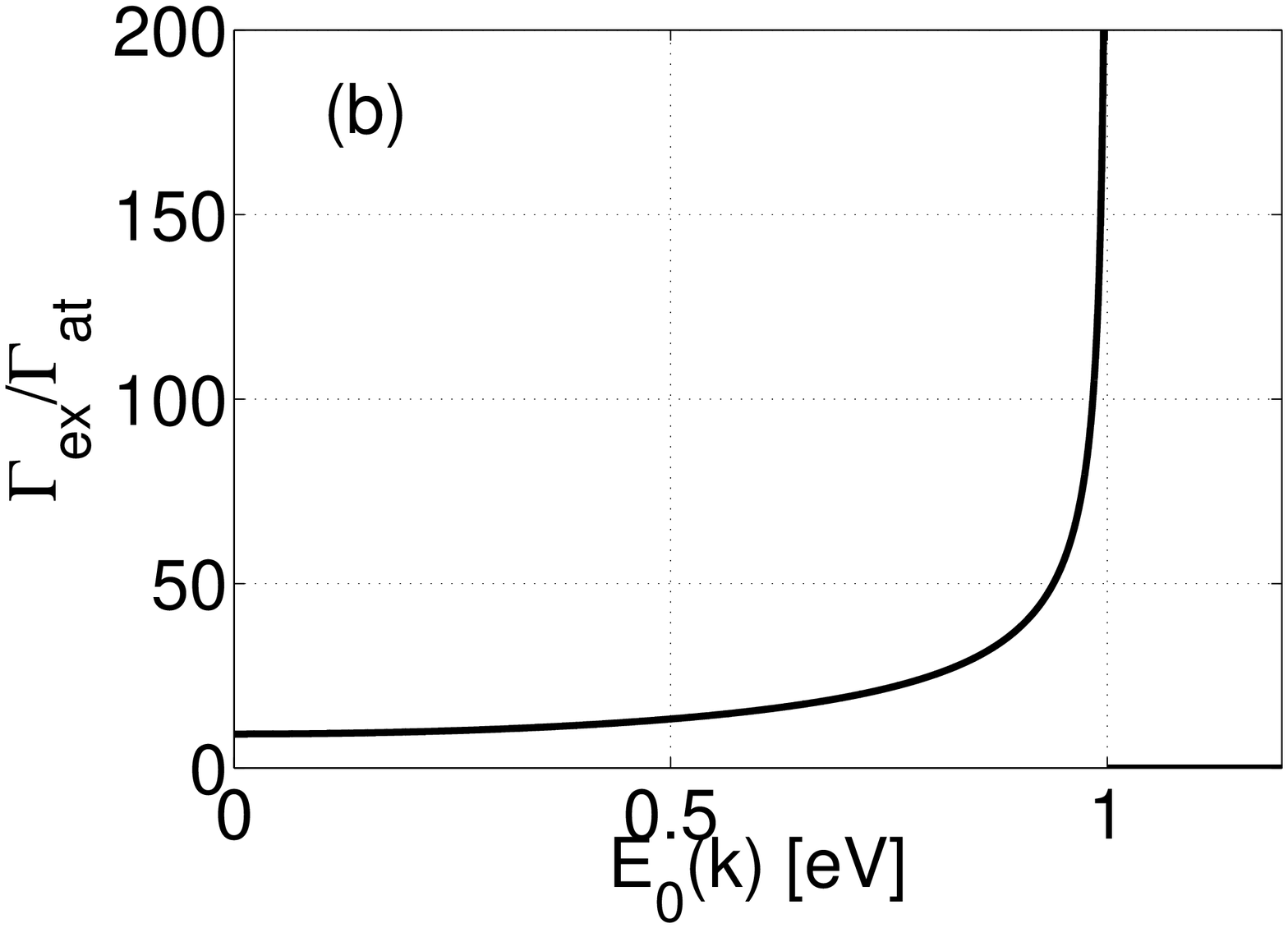}}
\caption{The scaled damping rate $\Gamma_{ex}/\Gamma_{at}$ vs. $E_{0}(k)=\hbar ck$, for $\theta=\pi/4$ with (a) $\phi=0$, and (b) $\phi=\pi/2$.}
\end{figure}

In figure (4) we plot $\Gamma_{ex}/\Gamma_{at}$ vs. $E_{0}(k)$ for $\theta=\pi/2$, where the transition dipole has only in-plane component. We take the in-plane wave vector ${\bf k}$ to be parallel to the in-plane transition dipole, that is $\phi=0$. For small wave vectors the excitons are superradiant with $\Gamma_{ex}$ larger than $\Gamma_{at}$. The damping rate decreases with increasing $k$ to become metastable, and vanishes for $k\geq k_c$.

\begin{figure}[h!]
\centerline{\epsfxsize=4cm \epsfbox{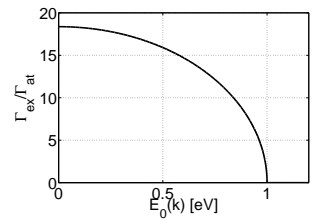}}
\caption{The scaled damping rate $\Gamma_{ex}/\Gamma_{at}$ vs. $E_{0}(k)=\hbar ck$, for $\theta=\pi/2$ and $\phi=0$.}
\end{figure}

In figures (5) we plot $\Gamma_{ex}/\Gamma_{at}$ vs. $\phi$ for $\theta=\pi/4$, where the transition dipole has equal in-plane and normal components. Figure (5a) is for small wave vectors $E_0=0.1\ eV$. Here the excitons are superradiant with maximum damping rate at $\phi=\pi$. In figure (5b) we plot for $E_0=0.9\ eV$.

\begin{figure}[h!]
\centerline{\epsfxsize=4cm \epsfbox{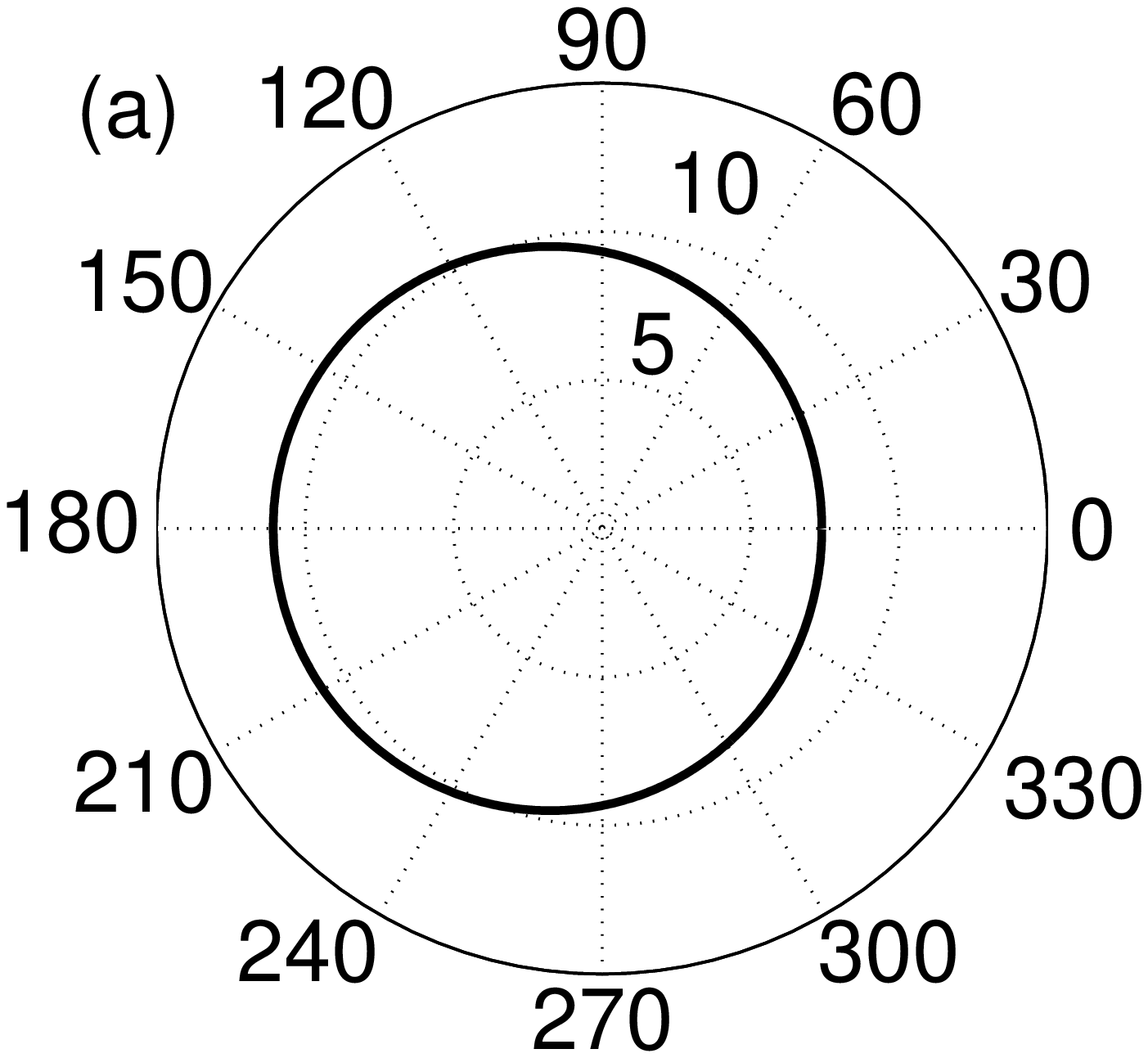}\epsfxsize=4cm \epsfbox{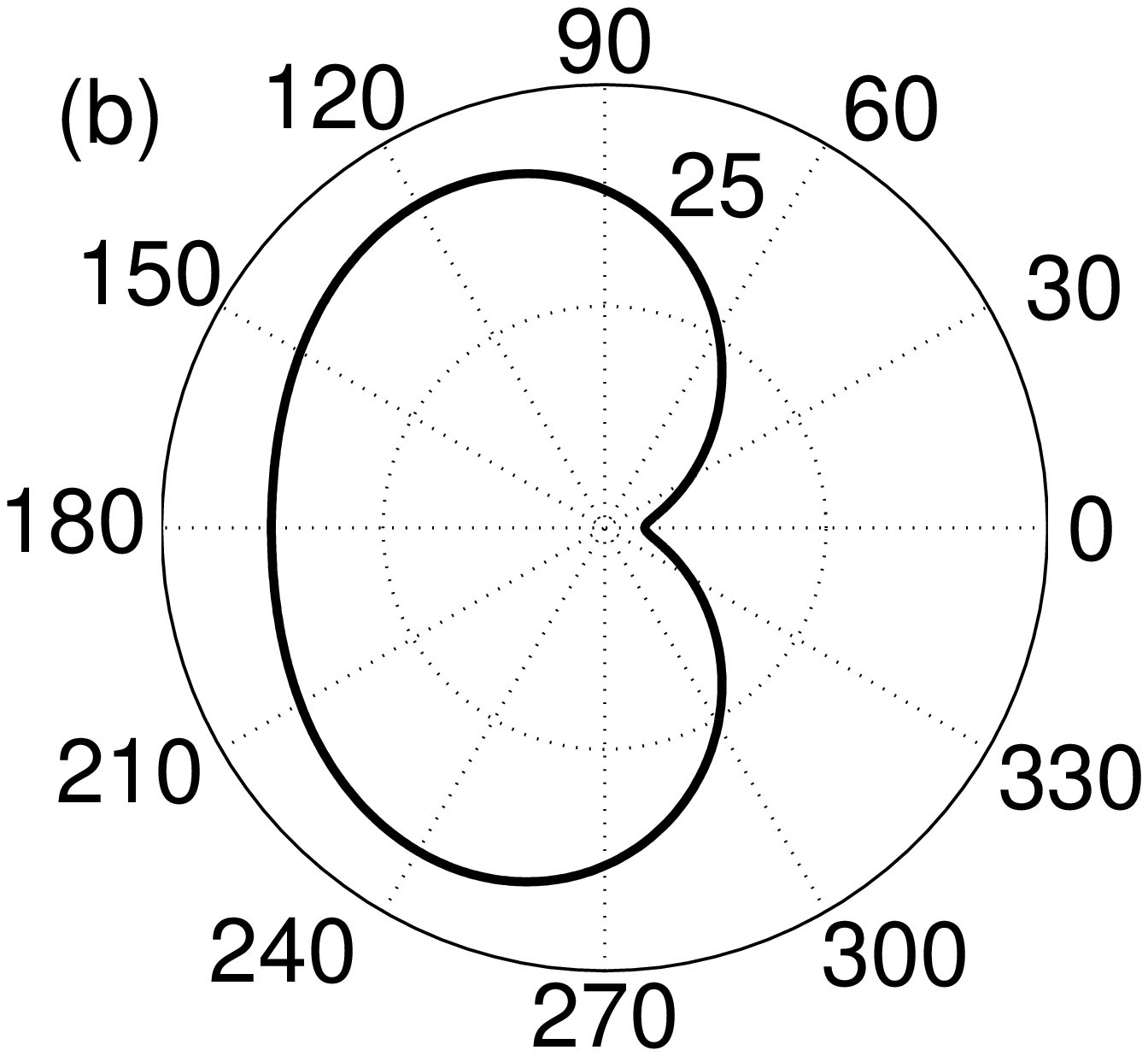}}
\caption{The scaled damping rate $\Gamma_{ex}/\Gamma_{at}$ vs. $\phi$, for $\theta=\pi/4$ with (a) $E_0=0.1\ eV$, and (b) $E_0=0.9\ eV$.}
\end{figure}

In figures (6) we plot $\Gamma_{ex}/\Gamma_{at}$ vs. $\phi$ for $\theta=\pi/2$, where the transition dipole has only in-plane component. Figure (6a) is for small wave vectors $E_0=0.1\ eV$ and figure (6b) for $E_0=0.9\ eV$. The excitons are superradiant.

\begin{figure}[h!]
\centerline{\epsfxsize=4cm \epsfbox{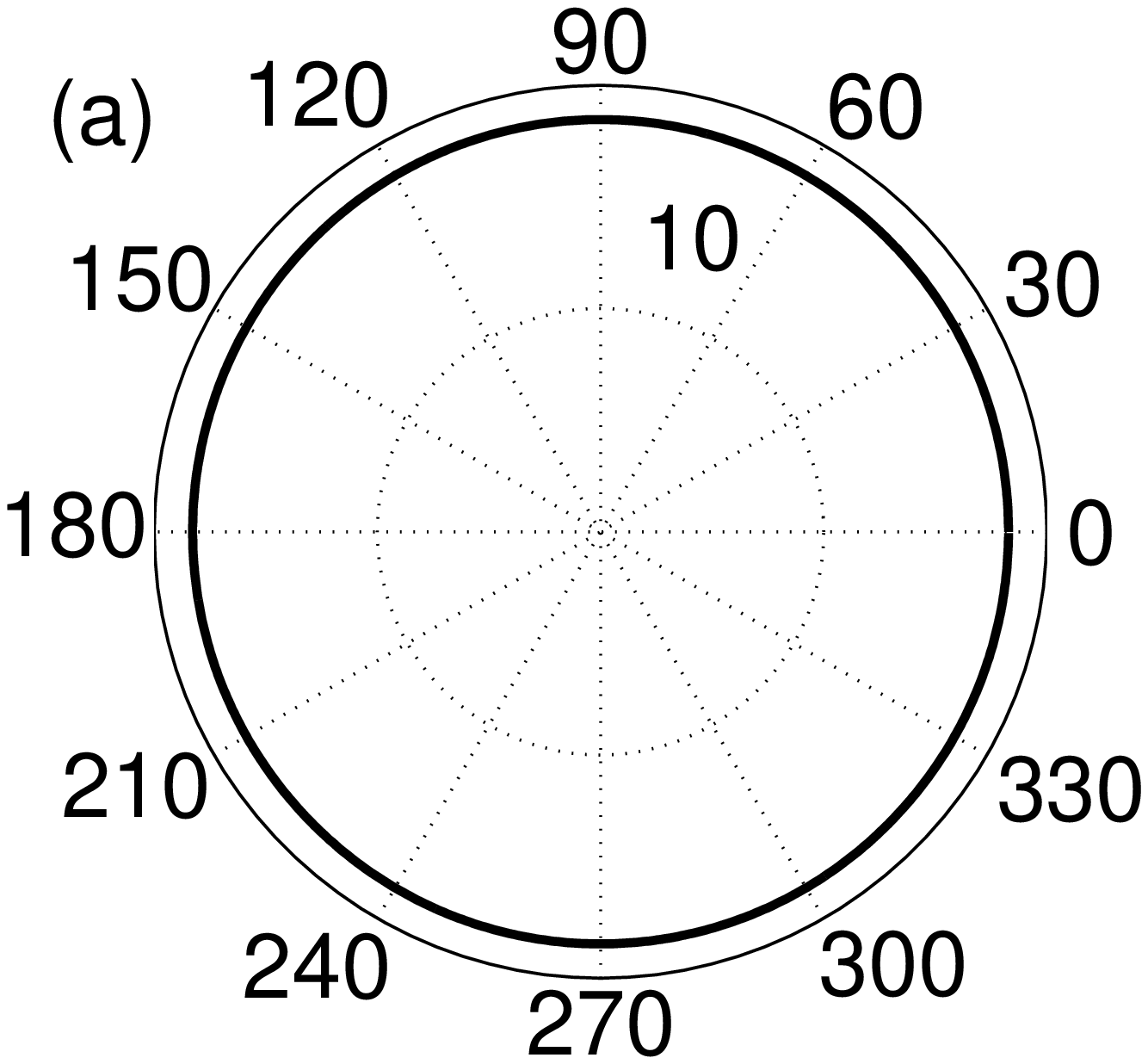}\epsfxsize=4cm \epsfbox{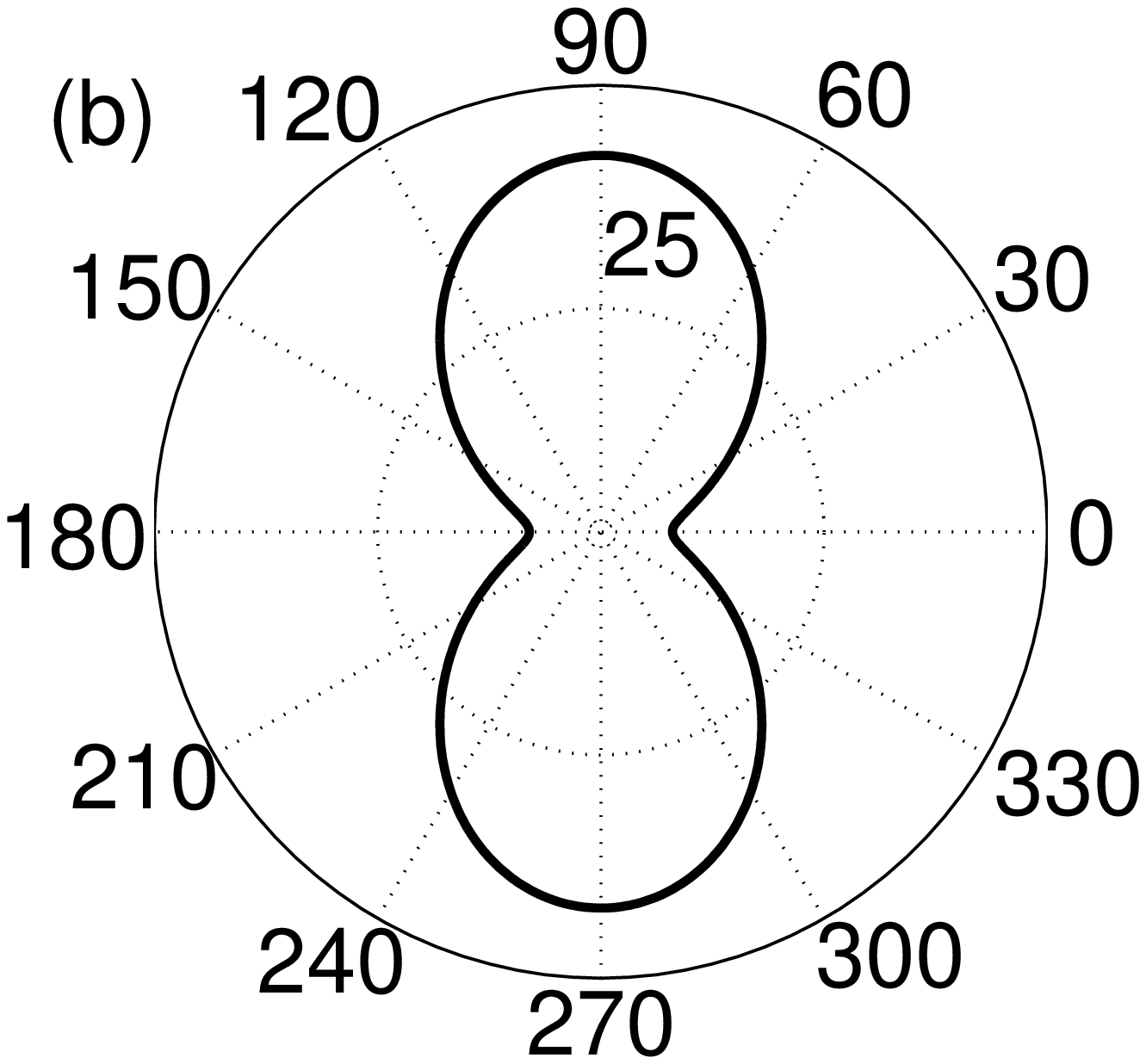}}
\caption{The scaled damping rate $\Gamma_{ex}/\Gamma_{at}$ vs. $\phi$, for $\theta=\pi/2$ with (a) $E_0=0.1\ eV$, and (b) $E_0=0.9\ eV$.}
\end{figure}

In figures (7) we plot $\Gamma_{ex}/\Gamma_{at}$ vs. $\theta$. In figure (7a) we use $\phi=\pi/2$ and $E_0=0.1\ eV$, and in figure (7b) we use $\phi=\pi/2$ and $E_0=0.9\ eV$, where here the excitons are superradiant with maximum damping rate at $\theta=\pi/2$.

\begin{figure}[h!]
\centerline{\epsfxsize=4cm \epsfbox{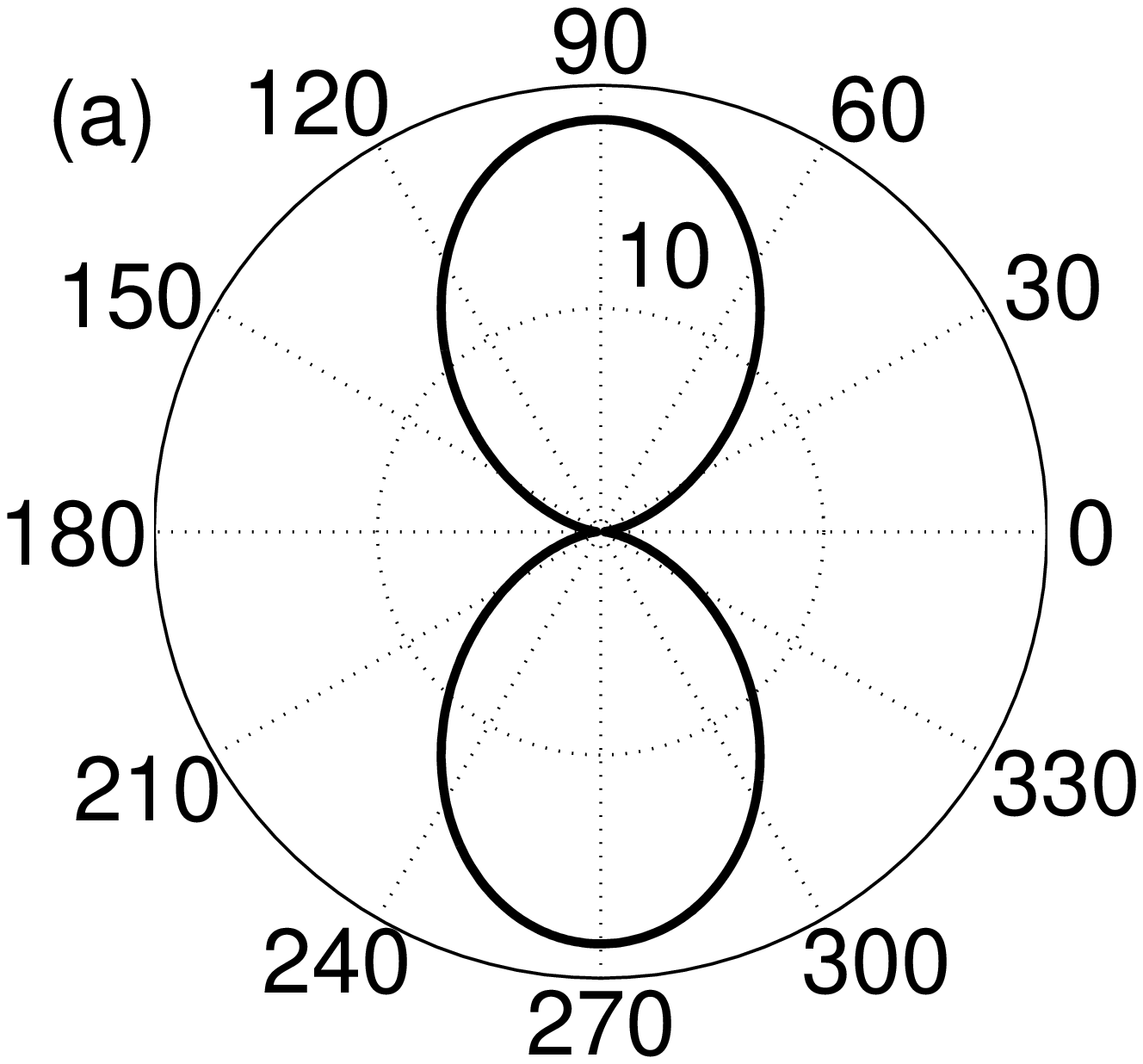}\epsfxsize=4cm \epsfbox{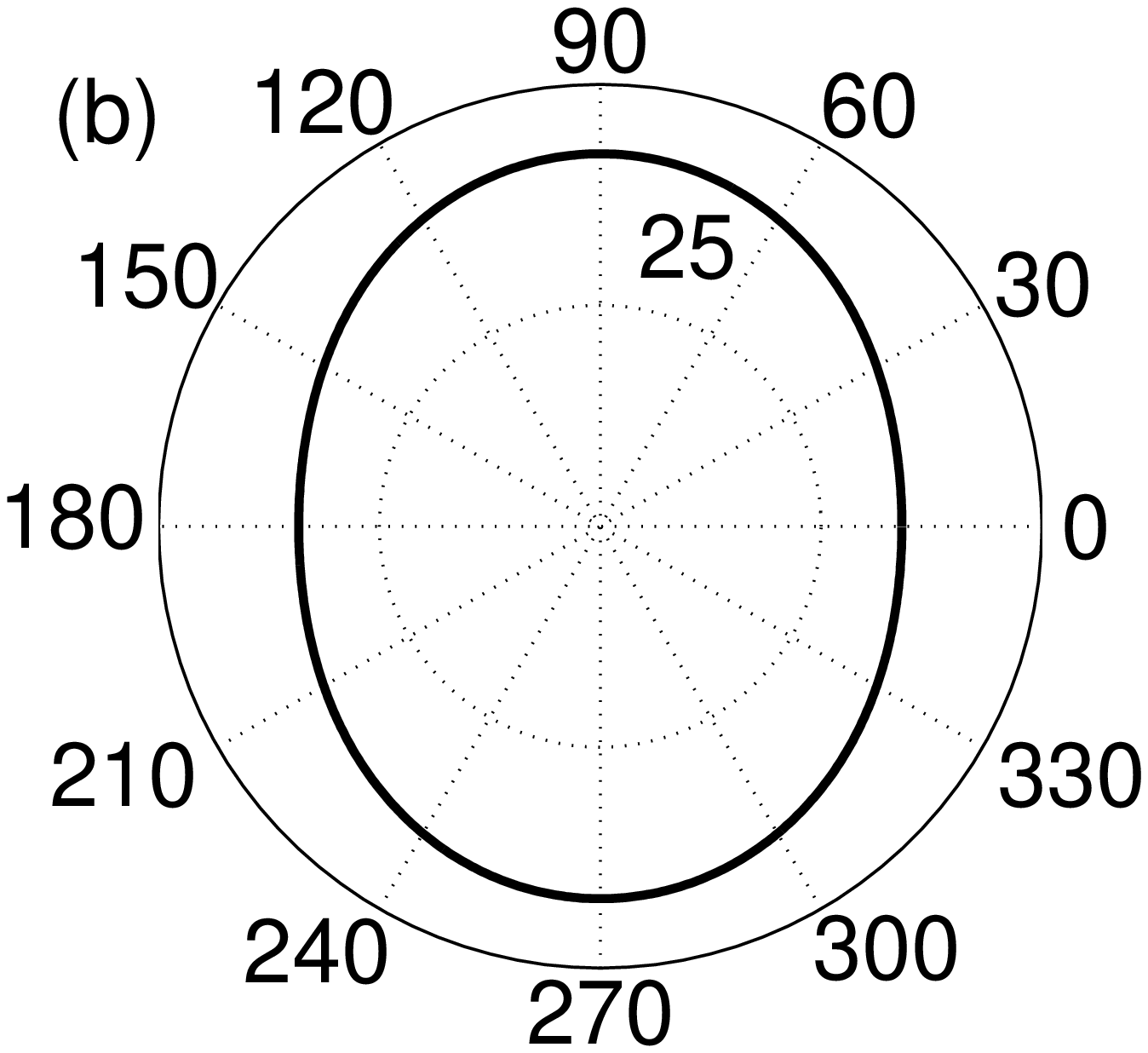}}
\caption{The scaled damping rate $\Gamma_{ex}/\Gamma_{at}$ vs. $\theta$, for $\phi=\pi/2$ with (a) $E_0=0.1\ eV$, and (b) $E_0=0.9\ eV$.}
\end{figure}

Now we calculate the emission pattern for a long wave length exciton of wave vector ${\bf k}$ into free space. Using the Hamiltonian $H=H_{ex}+H_{rad}+H_I$, we derive the radiation field operator equation of motion $i\frac{d\ b_{{\bf q}\lambda}(t)}{dt}=\omega_r(q)\ b_{{\bf q}\lambda}(t)-ig_{{\bf q}\lambda}\ B_{\bf k}$, which has the formal solution
\begin{equation}
b_{{\bf q}\lambda}(t)=b_{{\bf q}\lambda}(0)\ e^{-i\omega_r(q)t}-g_{{\bf q}\lambda}\int_0^{t}dt'\ B_{\bf k}(t')\ e^{-i\omega_r(q)(t-t')}.
\end{equation}
In using the source term, the positive part of the electric field for a fixed ${\bf k}$, at position ${\bf r}$ and time $t$, is
\begin{eqnarray}
&&\hat{\bf E}^+_{rad}({\bf r},t)=i\sum_{q_z\lambda}\frac{\omega_r(q)\sqrt{N}}{2\epsilon_0 V}\ {\bf e}_{{\bf q}\lambda}\left(\mbox{\boldmath$\mu$}\cdot{\bf e}_{{\bf q}\lambda}\right)\ e^{i\left[{\bf k}\cdot\mbox{\boldmath$\rho$}+q_zz\right]} \nonumber \\
&\times&e^{-i\omega_e({\bf k})t}\int_0^{t}dt'\ \tilde{B}_{\bf k}(t')\ e^{i\left[\omega_e({\bf k})-\omega_r(q)\right](t-t')},
\end{eqnarray}
with ${\bf r}=\mbox{\boldmath$\rho$}+z{\bf z}$, where we transferred the exciton operator into rotating frame by using $B_{\bf k}(t)=\tilde{B}_{\bf k}(t)\ e^{-i\omega_e({\bf k})t}$, with $E_{ex}({\bf k})=\hbar\omega_e({\bf k})$. The sum over $q_z$ casts as before into an integral, and we use also the summation over the photon polarization $\sum_{\lambda}{\bf e}_{{\bf q}\lambda}{\bf e}_{{\bf q}\lambda}=1-\frac{{\bf q}{\bf q}}{q^2}$.

At this point we take the approximation of long wave length excitons, that is small wave vectors with $ka\ll 1$, where $k\ll q_z$ to obtain
\begin{eqnarray}
&&\hat{\bf E}^+_{rad}({\bf r},t)\simeq i\frac{c}{4\pi\epsilon_0a^2\sqrt{N}}\int_0^{\infty}dq_z q_z\left(\mbox{\boldmath$\mu$}-\frac{{\bf q}\left({\bf q}\cdot\mbox{\boldmath$\mu$}\right)}{q_z^2}\right) \nonumber \\
&\times&e^{i\left[{\bf k}\cdot\mbox{\boldmath$\rho$}+q_zz-\omega_e({\bf k})t\right]}\int_0^{t}dt'\ \tilde{B}_{\bf k}(t')\ e^{i\left[\omega_e({\bf k})-cq_z\right](t-t')}.
\end{eqnarray}

We use $\mbox{\boldmath$\mu$}=(\mbox{\boldmath$\mu$}_{\parallel},\mu_z)$, and ${\bf q}=({\bf k},q_z)$, and substitute $\omega=cq_z$ and $d\omega=c\ dq_z$. Then we apply the Weisskopf-Wigner Approximation \cite{Wolf}, in replacing $\omega$ under the integrand by $\omega_e({\bf k})$ and to take them out of the integral, and to extend the lowest limit of the integral over $\omega$ to $-\infty$. After applying the result $\int_{-\infty}^{+\infty}d\omega\ e^{i\left[\omega_e({\bf k})-\omega\right]\left(t-t'-\frac{z}{c}\right)}=2\pi\delta\left(t-t'-\frac{z}{c}\right)$, in terms of the angle $\phi$ and in the laboratory frame, we have
\begin{eqnarray}
\hat{\bf E}^+_{rad}({\bf r},t)&\simeq& i\frac{E_0(k)}{2\epsilon_0\hbar ca^2\sqrt{N}}e^{i{\bf k}\cdot\mbox{\boldmath$\rho$}}B_{\bf k}(t-z/c)\left\{\frac{E_{ex}({\bf k})}{E_0(k)}\mbox{\boldmath$\mu$}_{\parallel}\right. \nonumber \\
&-&\left.\left[\frac{E_0(k)}{E_{ex}({\bf k})}\mu_{\parallel}\cos\phi+\mu_z\right]\hat{\bf k}\ ,\ -\mu_{\parallel}\cos\phi\right\},
\end{eqnarray}
where we defined the unit vector $\hat{\bf k}={\bf k}/k$.

The expectation value for the exciton operators is
\begin{equation}
\langle B_{\bf k}^{\dagger}(t-z/c)B_{\bf k}(t-z/c)\rangle=\langle B_{\bf k}^{\dagger}(0)B_{\bf k}(0)\rangle\ e^{-\Gamma_{\bf k}(t-z/c)},
\end{equation}
where $\Gamma_{\bf k}$ is the previous exciton damping rate.

For a single exciton, where $\langle B_{\bf k}^{\dagger}(0)B_{\bf k}(0)\rangle=1$, and in terms of the angle $\theta$, we obtain the emission intensity
\begin{eqnarray}
&&\langle\hat{\bf E}^-_{rad}\hat{\bf E}^+_{rad}\rangle_{\bf k}=\frac{E_0^2(k)\mu^2}{(2\epsilon_0\hbar ca^2)^2N}\left\{\cos^2\theta+\sin^2\theta\left[\frac{E_{ex}^2({\bf k})}{E_0^2(k)}\right.\right. \nonumber \\
&&\left.\left.+\cos^2\phi\left(\frac{E_{0}^2(k)-E_{ex}^2({\bf k})}{E_{ex}^2({\bf k})}\right)\right]+2\sin\theta\cos\theta\cos\phi\right. \nonumber \\
&&\left.\times\left(\frac{E_{0}^2(k)-E_{ex}^2({\bf k})}{E_{ex}({\bf k})E_{0}(k)}\right)\right\}\ e^{-\Gamma_{\bf k}(t-z/c)}.
\end{eqnarray}
The dependence on the distance from the lattice plane appears only in the damping exponent. Note that the result is in the limit of $ka\ll 1$.

\ 

In summary our calculations predict that the radiative damping rate and emission pattern into free space of an exciton, for two dimensional optical lattice ultracold atoms in the Mott insulator phase, is strongly deviated from a single excited atom. The emitted light properties are a complicated function of the exciton wave number and polarization direction. For some wave numbers and polarizations excitons exhibit damping rate much larger than a single excited atom and can be considered as superradiant excitons, while for others the damping rate is much smaller. We found that beyond a critical wave number excitons become metastable with zero damping rate, where beyond $\hbar ck_c=E_{ex}({\bf k}_c)$ no free space photon can conserve both momentum and energy with an exciton with a fixed wave vector. Metastable excitons can store photons for a long time and transport them over a long range. 

The big challenge is how to excite metastable excitons in such a system. As they decouple from the  free space radiation field, they can not be excited directly by uniformly shining a light on the optical lattice.  However, by localized excitation or targeting the edge of the lattice we can excite at least a partial amplitude in the metastable domain, which should survive much longer. As an alternative one could use Raman excitation in a properly phase matched geometry to excite long lived excitons, which can be read out by a Stokes pulse. Note that some metastable excitons are bright states, and hence by using different optical elements one can excite them directly, or for a finite system to shine the optical lattice from the side. Excitons are collective electronic excitations where their momentum is distributed over all the lattice sites and not on a single atom site, therefore shining light directly on the optical lattice in order to excite excitons will not destroy the Mott insulator phase. The results are of importance for quantum optics and quantum information processing, and are valid for any planar periodic structure of optically active materials.

\ 

The work was supported by the Austrian Science Funds (FWF), via the project (P21101).

\end{document}